\documentclass[12pt,english,floatfix,showkeys,superscriptaddress,aps,prd,preprint]{revtex4}
\usepackage[latin1]{inputenc}
\usepackage[T1]{fontenc}
\usepackage{lmodern}
\usepackage{amsmath}
\usepackage{amssymb}
\usepackage{graphicx}
\usepackage{float}
\usepackage{esint}
\usepackage{longtable}
\usepackage{dcolumn}
\usepackage{babel}
\usepackage{csquotes}
\usepackage{color}
\usepackage{slashed}
\usepackage{simplewick}
\usepackage{amsmath,latexsym}

\usepackage{hyperref}
\hypersetup{
    colorlinks,
    citecolor=blue,
    filecolor=green,
    linkcolor=purple,
    urlcolor=red,
}

\usepackage{slashed}

\usepackage{hyperref}
\hypersetup{colorlinks,breaklinks,
			citecolor=[rgb]{0,0.0,1.0},
            urlcolor=[rgb]{0.0,0.0,1.0},
            linkcolor=[rgb]{0,0.5,0.9}}

\begin{document}

\title{Localization of abelian gauge fields with Stueckelberg-like geometrical coupling on $f(T,B)$-thick brane}

\author{F. M. Belchior}
\email{belchior@fisica.ufc.br}
\affiliation{Universidade Federal do Cear\'a (UFC), Departamento de F\'isica,\\ Campus do Pici, Fortaleza - CE, C.P. 6030, 60455-760 - Brazil.}
\author{A. R. P. Moreira}
\email{allan.moreira@fisica.ufc.br}
\affiliation{Universidade Federal do Cear\'a (UFC), Departamento de F\'isica,\\ Campus do Pici, Fortaleza - CE, C.P. 6030, 60455-760 - Brazil.}
\author{R. V. Maluf}
\email{r.v.maluf@fisica.ufc.br}
\affiliation{Universidade Federal do Cear\'a (UFC), Departamento de F\'isica,\\ Campus do Pici, Fortaleza - CE, C.P. 6030, 60455-760 - Brazil.}
\affiliation{Departamento de F\'{i}sica Te\'{o}rica and IFIC, Centro Mixto Universidad de Valencia - CSIC. Universidad
de Valencia, Burjassot-46100, Valencia, Spain.}

\author{C. A. S. Almeida}
\email{carlos@fisica.ufc.br}
\affiliation{Universidade Federal do Cear\'a (UFC), Departamento de F\'isica,\\ Campus do Pici, Fortaleza - CE, C.P. 6030, 60455-760 - Brazil.}


\date{\today}

\begin{abstract}
In the context of $f(T,B)$ modified teleparallel gravity, we investigate the influence of torsion scalar $T$ and boundary term $B$ on the confinement of both the gauge vector and Kalb-Ramond fields. Both fields require a suitable coupling in five-dimensional braneworld scenarios to yield a normalizable zero mode. We propose a Stueckelberg-like geometrical coupling that non-minimally couples the fields to the torsion scalar and boundary term. To set up our braneworld models, we use the first-order formalism in which two kinds of superpotential are taken: sine-Gordon and $\phi^{4}$-deformed. The geometrical coupling is used to produce a localized zero mode. Moreover, we analyze the massive spectrum for both fields and obtain possible resonant massive modes. Furthermore, we do not find tachyonic modes leading to a consistent thick brane.
\end{abstract}


\maketitle


\section{Introduction}

Since its formulation by Albert Einstein in 1916, general relativity (GR) has successfully described a wide range of astrophysical phenomena. However, there remain several outstanding questions, such as matter-antimatter asymmetry \cite{Dine:2003ax}, dark matter \cite{darkmatter}, dark energy \cite{{Peebles:2002gy}}, inflation \cite{Nojiri:2017ncd}, cosmological constant \cite{cosmologicalconstant}, and the hierarchy problem \cite{rs2,rs}, which have motivated investigations into modified gravity theories. Among the simplest extensions of GR are $f(R)$ \cite{DeFelice:2010aj, Bisabr:2010xy,Bazeia:2014poa} and $f(R,\mathcal{T})$ \cite{Deb:2017rhc, Moraes:2016akv,Bazeia:2021bwg,Gu:2014ssa} gravity theories, which have garnered attention as possible solutions to these open problems.

Constructing a nontrivial curvature-free spacetime has been explored within the context of modified gravity theories, with the teleparallel equivalent of general relativity (TEGR)  receiving particular attention \cite{Hayashi1979, deAndrade1997, deAndrade1999, Aldrovandi,Maluf:2013gaa,Bahamonde:2021gfp}. In this theory, the tetrad (vielbein) is used as the dynamic variable instead of the metric, and the gravitational action is defined in terms of the torsion scalar $T$. Analogous to the $f(R)$ model for GR, the $f(T)$ gravity \cite{Ferraro2011us, Tamanini2012,Yang2017,ftenergyconditions} represents a natural modification of TEGR, with the main difference being that $f(R)$ generates fourth-order equations while $f(T)$ generates second-order equations. Recently, other modified gravity models such as $f(T,B)$ and $f(T,\mathcal{T})$ gravities have been proposed and investigated in various contexts, including cosmology, black holes, wormholes, and braneworld scenarios \cite{Franco2020,EscamillaRivera2019,Bahamonde2015, Wright2016, Bahamonde2016,Bahamonde2016a, Caruana2020, 
Arouko2020g,Ghosh2020,Salako2020t,Mirzaei2020, Pourbagher2020,Bahamonde2020a,Azhar2020,Bhattacharjee2020,Moreira:2021xfe,Moreira:2021vcf,Moreira:2021cta,Moreira:2021uod}.

Concerning the possibility of extra dimensions, Lisa Randall and Raman Sundrum proposed two groundbreaking papers \cite{rs,rs2} that presented a solution to the hierarchy problem by embedding our Universe as a brane in a warped five-dimensional spacetime. They argued that gravity is free to propagate throughout the bulk. At the same time, the standard model fields are confined to the brane, explaining the observed weakness of gravity in our four-dimensional spacetime. Since then, braneworld scenarios have been extensively investigated \cite{Gremm1999}. In this sense, thick brane models \cite{Bazeia:2003aw, Bazeia:2004dh, Bazeia:2005hu, Bazeia:2021jok, Afonso2006, Cruz:2018qby, Janssen2007} are a direct extension of the Randall-Sundrum models and are supported by the addition of real scalar fields that induce an internal structure \cite{Bazeia:1995ui}. Many thick brane models have been proposed in various contexts, including the Bloch brane \cite{ Chumbes:2011zt, Cruz:2010zz, Cruz:2012kd, Cruz:2013zka}, cuscuton braneworld \cite{Bazeia:2021jok,Bazeia:2021bwg} and modified gravity such as $f(R)$ gravity \cite{Gu:2014ssa,Bazeia:2014poa}, $f(R,T)$ gravity \cite{Bazeia:2021bwg, ftborninfeld}, Gauss-Bonnet and teleparallel gravity \cite{Menezes, Belchior}. In the braneworld context, it is crucial to investigate gravity localization and analyze how standard model fields, including gauge fields \cite{Cruz:2015nrd} and fermions \cite{Liu2009}, are confined to the brane. Generally, a mechanism is required to create a normalizable zero mode, which can be achieved through non-trivial couplings that typically depend on the extra dimension. For example, in Ref. \cite{Chumbes:2011zt}, it was considered a non-minimal coupling between the gauge and Kalb-Ramond fields and the background scalar field directly in the kinetic term. In Refs. \cite{Cruz:2010zz, Cruz:2012kd, Cruz:2013zka}, a similar coupling was employed, but considering the dilaton field. 

As demonstrated by several studies, geometric couplings have been an essential mechanism for localizing gauge fields \cite{Zhao:2014gka, Alencar:2014moa, Alencar:2015oka, Alencar:2015awa, Vaquera-Araujo:2014tia, Zhao:2014iqa, Zhao:2017epp}. These couplings can be introduced in either the kinetic or massive term of the field equation, leading to a normalizable zero mode. However, in field theory, a mass term breaks the gauge symmetry. To overcome this difficulty, we can utilize the Stueckelberg mechanism \cite{Stueckelberg:1938hvi,Ruegg:2003ps}. This procedure introduces an auxiliary scalar field to restore the gauge symmetry. Therefore, it would be interesting to extend this investigation into a five-dimensional warped spacetime \cite{Vaquera-Araujo:2014tia,Zhao:2014iqa,Zhao:2017epp}.

This work studies the localization of gauge vector and Kalb-Ramond fields in $f(T,B)$-braneworld scenarios. Geometrical coupling is achieved through a Stueckelberg-like term, non-minimally coupling the fields to the torsion scalar and the boundary term. The analysis employs supersymmetric quantum mechanics and demonstrates the absence of tachyon modes, which are crucial for the consistency of the physical system \cite{Skenderis:1999mm,DeWolfe:1999cp}. The confinement of the zero mode for abelian gauge fields is also demonstrated, complementing previous studies on gravity localization \cite{Moreira:2021xfe} and fermion localization \cite{Moreira:2021vcf} in similar thick brane models.

Our paper is organized as follows. In Sec. (\ref{sec2}), the main aspects of $f(T, B)$ gravity, as well as the braneworld scenarios constructed in such gravity, are briefly reviewed. In Sec. (\ref{sec3}), a Stueckelberg-like coupling is employed to localize both gauge vector and the Kalb-Ramond fields on thick brane generated by a single real scalar field in $f(T, B)$ gravity. Finally, our conclusions are discussed in Sec. (\ref{sec4}).

\section{Thick brane in $f(T,B)$ gravity}
\label{sec2}

Let us begin by briefly establishing our notation and conventions for braneworld models in the context of teleparallel gravity and its generalization to $f(T,B)$ gravity. 

The usual description of general relativity as a Riemannian spacetime involves a metric $g_{AB}$ as basic dynamical object and a covariant derivative $\nabla_{M}$ constructed with the Levi-Civita connection $\Gamma^{A}{}_{BC}=\frac{1}{2}g^{AD}(\partial_{B}g_{DC}+\partial_{C}g_{DB}-\partial_{D}g_{BC})$. This connection is torsion-free and obeys the metricity condition $\nabla_{M} g_{NP}=0$ by definition. On the other hand, in teleparallel theories of gravity, the fundamental dynamical variables are the tetrad fields (or vierbein) $h^{a}{}_{M}$, responsible for the conversion from the local Lorentz frame to spacetime coordinates through the relation
\begin{equation}
A_{MN\cdots}=h^{a}{}_{M}h^{b}{}_{N}\cdots A_{ab\cdots},
\end{equation}where $A_{MN\cdots}$ represents the covariant components of an arbitrary tensor field in a coordinate basis and $A_{ab\cdots}$ the corresponding covariant components in a local Lorentz frame. In this way, the tetrads and their inverse $h^{M}{}_{a}$ transform the metric and the inverse metric $g^{MN}$ to the Minkowski form $\eta_{ab}$ via
\begin{equation}
g_{MN}=h^{a}{}_{M}h^{b}{}_{N}\eta_{ab},\ \ \ \ \ g^{MN}=h^{M}{}_{a}h^{N}{}_{b}\eta^{ab},\label{g_eta}
\end{equation}where the bulk coordinate indices are denoted by capital Latin index $M=0,\ldots,D-1$, whereas the vielbein indices are denoted by Latin indices $a=0,\ldots,D-1$. Also, the Minkowski metric has the signature $(-,+,\cdots,+)$. For consistency, the tetrads and their inverse must obey the orthogonality relations
\begin{equation}
h^{a}{}_{M}h^{M}{}_{b}=\delta_{b}^{a},\ \ \ \ h^{M}{}_{a}h^{a}{}_{N}=\delta_{N}^{M}.
\end{equation}

In a curvature-free formulation of the teleparallel theory, we can define an object called the Weitzenb\"{o}ck connection by \cite{Aldrovandi,Maluf:2013gaa,Bahamonde2015}
\begin{equation}
\widetilde{\Gamma}^{P}{}_{MN}\equiv h^{P}{}_{a}\partial_{M}h^{a}{}_{N},
\end{equation}such that the condition of absolute parallelism, namely \cite{Hayashi1979}
\begin{equation}
\widetilde{\nabla}_{Q}h^{a}{}_{M}=\partial_{Q}h^{a}{}_{M}-\widetilde{\Gamma}^{P}{}_{QM}h^{a}{}_{P}=0,   
\end{equation}is consistently satisfied. It can be explicitly verified that the Riemann and Ricci tensors are identically zero when calculated with the Weitzenb\"{o}ck connection. In turn, we can define the associated torsion tensor as
\begin{equation}
T^{P}{}_{MN}=\widetilde{\Gamma}^{P}{}_{MN}-\widetilde{\Gamma}^{P}{}_{NM}.
\end{equation}The difference between the Weitzenb\"{o}ck and Levi-Civita connections define the contortion tensor, which can be expressed via torsion tensor as
\begin{equation}
K^{P}{}_{MN}=\frac{1}{2}(T^{P}{}_{MN}+T_{M}{}^{P}{}_{N}+T_{N}{}^{P}{}_{M}).
\end{equation}
The contortion tensor has two traces. One of them is identically null $K^{P}{}_{MP}=0$, while the other is related to the torsion trace vector $T_{M}\equiv T^{P}{}_{MP}$ by
\begin{equation}
K^{P}{}_{PM}=-K_{M}{}^{P}{}_{P}=-T_{M}.
\end{equation}
Another relevant object related to the contortion and torsion tensor is the superpotential tensor, which can be written as  
\begin{equation}
S_{P}{}^{MN}=\frac{1}{2}(K^{M}{}_{P}{}^{N}+\delta_{P}^{M}T^{N}-\delta_{P}^{N}T^{M}).
\end{equation}
The link between the GR and TEGR is established through the relation \cite{Maluf:2013gaa,Bahamonde2015}
\begin{equation}
R=-T-2\nabla^{M}T^{N}{}_{MN},\label{RTB}
\end{equation}where the Ricci scalar is calculated in terms of the Levi-Civita connection, and the scalar $T$, called the torsion scalar, is given by
\begin{equation}
T=S_{P}{}^{MN}T^{P}{}_{MN}.\label{Ts}
\end{equation}
From Eq. (\ref{RTB}), one can identify the boundary term as
\begin{equation}
B\equiv-2\nabla^{M}T^{N}{}_{MN}=-\frac{2}{h}\partial_{M}(hT^{M}),\label{Bs}
\end{equation}where $h=\sqrt{-g}$, with $g$ being the determinant of the metric tensor.

Hence, one can see that GR and TEGR are two equivalent theories of gravity since the Ricci scalar and the torsion scalar are related by a boundary term, namely $R=-T+B$. However, this equivalence does not hold in more general theories of gravity, like $f(R)$ and $f(T)$ gravity, which lead to modified equations of motion \cite{Bahamonde2015,Bahamonde:2021gfp,Caruana2020}. 

A general class of modified teleparallel gravity can be constructed by using an arbitrary function of $T$ and $B$, resulting in what is known as $f(T,B)$ gravity  \cite{Bahamonde2016,EscamillaRivera2019,Bahamonde2016a,Caruana2020,Azhar2020,Moreira:2021xfe,Moreira:2021cta}. In this framework, the gravitational action in a $D$-dimensional spacetime can conveniently be expressed as
\begin{equation}
S_{G}=\int d^{D}xh\left(\frac{1}{2\kappa_{D}}f(T,B)+\mathcal{L}_{m}\right),\label{Action1}
\end{equation}where $\kappa_{D}$ is a suitable constant with mass dimension 
$[\kappa_{D}]=M^{2-D}$ (in natural units), and $\mathcal{L}_m$ is the matter Lagrangian which will be introduced later. 

The gravitational field equations follow from the action (\ref{Action1}) by varying with respect to the vierbein (See Ref. \cite{Bahamonde2015} and references therein for a detailed derivation). Explicitly, we find
\begin{eqnarray}
&& h^{M}{}_{a}\Box f_{B}-h^{P}{}_{a}\nabla^{M}\nabla_{P}f_{B}+\frac{1}{2}h^{M}{}_{a}Bf_{B}+2\Big[(\partial_{P}f_{B})+(\partial_{P}f_{T})\Big]S_{a}{}^{PM}\nonumber\\
&& +\frac{2}{h}\partial_{P}(hS_{a}{}^{PM})f_{T}-2f_{T}T^{P}{}_{Na}S_{P}{}^{MN}-\frac{1}{2}h^{M}{}_{a}f=\kappa_{D}\mathcal{T}^{M}{}_{a},\label{EoM1}
\end{eqnarray}where $\square=\nabla_M\nabla^M$, and we have defined $f\equiv f(T,B)$, $f_{T}\equiv \partial f(T,B)/\partial T$, $f_{B}\equiv \partial f(T,B)/\partial B$.  In the above equation the energy-momentum tensor is defined as
\begin{equation}
\mathcal{T}^{M}{}_{a}=\frac{1}{h}\frac{\delta( h \mathcal{L}_{m})}{\delta h^{a}\ _{M}}.
\end{equation}

For our present purposes, it is convenient to write the equations of motion in terms of spacetime indices only. It can be shown that Eq. (\ref{EoM1}) takes the following covariant form \cite{Bahamonde2015,Bahamonde2020a}:
\begin{eqnarray}
&&-f_{T}G_{MN}+(g_{MN}\Box-\nabla_{M}\nabla_{N})f_{B}+\frac{1}{2}(Bf_{B}+Tf_{T}-f)g_{MN}\nonumber\\
&&+2\Big[(f_{BB}+f_{BT})(\nabla_{P}B)+(f_{TT}+f_{BT})(\nabla_{P}T)\Big]S_{N}{}^{P}{}_{M}=\kappa_{D}\mathcal{T}_{MN},\label{EqMovMetric}
\end{eqnarray}where $G_{MN}\equiv R_{MN}-\frac{1}{2} g_{MN}R$ is the Einstein tensor calculated
with the Levi-Civita connection and $\mathcal{T}_{MN}\equiv h^{a}\ _{N}\mathcal{T}_{Ma}$ is the standard energy-momentum tensor.

After defining the main concepts of $f(T,B)$ gravity, we proceed to develop thick brane scenarios to investigate the localization of gauge fields, as detailed in the following section. In particular, we consider a warped five-dimensional spacetime where the metric is given by \cite{rs,rs2,Gremm1999,Bazeia:2003aw, Bazeia:2004dh, Bazeia:2005hu, Bazeia:2021jok, Afonso2006, Cruz:2018qby, Janssen2007}
\begin{equation}
ds^2=e^{2A(y)}\eta_{\mu\nu}dx^\mu dx^\nu + dy^2,\label{warp}
\end{equation} with $\eta_{\mu\nu}=\mbox{Diag}(-1,1,1,1)$ being the Minkowski spacetime metric, and the indices $\mu$, $\nu$ run over 0 to 3. The function $A(y)$ characterizes the warp factor and will be fixed with the particular potential choice. 

Now, let us take the \textit{funfbein} (vielbein defined in the five-dimensional spacetime) as 
\begin{equation}\label{000098}
h^{a}{}_{M}=\begin{pmatrix}
e^{A}\delta^\mu_\nu & 0\\
0 & 1
\end{pmatrix},
\end{equation}which ensures that the warped metric (\ref{warp}) is reproduced via relations (\ref{g_eta}). For this choice, we can determine the torsion scalar and the boundary term through Eqs. (\ref{Ts}-\ref{Bs}) as
\begin{eqnarray}
T&=&-12A'^{2},\nonumber\\
B&=&-8A''-32A'^{2},
\end{eqnarray}where prime $(')$ stands for derivative concerning the extra-coordinate $y$.

Consistent with many braneworld scenarios proposed in the literature \cite{Moreira:2021xfe,Moreira:2021vcf,Moreira:2021cta,Moreira:2021uod,Belchior,Cruz:2015nrd}
, we assume that a single scalar field generates the thick brane and has a standard Lagrangian given by
\begin{equation}
\mathcal{L}_{m}=-\frac{1}{2}\partial_{M}\phi\partial^{M}\phi-V(\phi),
\label{l11}
\end{equation}where $\phi\equiv \phi(y)$ depends only on the extra dimension $y$. 
Then, this choice leads to the following equation of motion for the matter field
\begin{equation}
g^{MN}\nabla_{M}\nabla_{N}\phi-\frac{\partial V}{\partial \phi}=0,\label{Eqmovphi}
\end{equation}with the associated energy-momentum tensor as
\begin{equation}
\mathcal{T}_{MN}=\partial_{M}\phi\partial_{N}\phi-\frac{1}{2}g_{MN}(2V+\partial_{P}\phi\partial^{p}\phi).
\end{equation}

By evaluating the field equations (\ref{EqMovMetric}) and (\ref{Eqmovphi}) under the conditions specified above, we obtain the following equations:
\begin{eqnarray}
&&-\frac{1}{2}f-\left(4A'^{2}+A''\right)\left(4f_{B}+3f_{T}\right)+f_{B}^{''}+3A'f_{B}'+24A'A'''\left(f_{BB}+f_{BT}\right)\nonumber\\
&&+24A'^{2}A''\left(8f_{BB}+11f_{BT}+3f_{TT}\right)=-\kappa_{5}V-\frac{1}{2}\kappa_{5}\phi'{}^{2},\label{w.2}
\end{eqnarray}
\begin{equation}
-\frac{1}{2}f-4A'^{2}\left(4f_{B}+3f_{T}\right)+4A'f_{B}'-4A''f_{B}=-\kappa_{5}V+\frac{1}{2}\kappa_{5}\phi'^{2},\label{w.1}
\end{equation}
\begin{equation}
\phi''+4A'\phi'=\frac{\partial V}{\partial\phi}.\label{w.0}
\end{equation}

After obtaining the gravitational field equations in the thick brane scenario, the next step is to specify the form of the $f(T,B)$ function. Since our purpose is to modify general relativity by considering the presence of torsion and an additional boundary term, we will limit ourselves to the most straightforward modification of TEGR, namely, a linear function given by $f(T,B)=c_1T+c_2B$. Here, $c_{1}$ and $c_{2}$ are the parameters that control the deviation from the usual gravity models, i.e., making $c_1=-1$ and $c_2=1$ we fall back into general relativity, but if we consider $c_1 =-1$ and $c_2=0$ we fall back into the usual teleparallelism.

\subsection{First-order formalism}

In order to find analytical solutions for our $f(T,B)$-braneworld system, as described by Eqs. (\ref{w.2})-(\ref{w.0}), we employ the Bogomolnyi-Prasad-Sommerfield (BPS) formalism \cite{Skenderis:1999mm, DeWolfe:1999cp}. This useful technique enables us to transform the second-order coupled field equations into first-order equations by utilizing auxiliary superpotentials \cite{Gremm1999,Afonso2006,Janssen2007}. In the context of modified teleparallel gravities, such a formalism has recently been used to find brane solutions in $f(T)$ \cite{Menezes,ftborninfeld} and $f(T,\mathcal{T})$ \cite{Moreira:2021uod} gravity models.

Considering $f(T,B)=c_1T+c_2B$, and fixing $\kappa_{5}=2$ for convenience, we can rewrite Eqs. (\ref{w.2}) and (\ref{w.1}) as 
\begin{eqnarray}
\label{e.10}
\frac{3}{2}c_1 A''&=& \phi'^2,\\
\label{e.20}
-3c_1A'^{2}&=&\frac{\phi'^2}{2}-V.
\end{eqnarray}
Note that the influence of the boundary term disappears, i.e., the parameter $c_2$ does not affect our solutions. This is expected for this scenario, as the same result occurs in usual teleparallelism, where the limiting term $B$ cancels out in the equations of motion.

To implement the first-order formalism, we introduce a superpotential $W(\phi)$ of the form
\begin{eqnarray}\label{09a}
A'=-\alpha W(\phi),
\end{eqnarray}where $\alpha$ is an arbitrary constant. Hence, the second-order equation (\ref{e.10}) turns into
\begin{eqnarray}\label{07}
\phi'=-\frac{3}{2}c_1\alpha W_\phi.
\end{eqnarray}
The potential $V$ is obtained through Eq. (\ref{e.20}), such that
\begin{eqnarray}\label{08}
V(\phi)= 3c_1\alpha^{2}W^{2}+\frac{9}{8}c_{1}^{2}\alpha^{2}W_{\phi}^{2}.
\end{eqnarray}

It is noteworthy that the default GR equations are restored when we consider $c_{1}=-1$ and $\alpha=\frac{1}{3}$ \cite{Gremm1999}, i.e., 
\begin{eqnarray}
\label{oo.3}\phi'&=&\frac{ W_\phi}{2},\\ 
\label{oo.4}V(\phi)&=&\frac{W_\phi ^2}{8}-\frac{W^2}{3}.
\end{eqnarray}

The energy density, $\rho(y)=-e^{2A(y)}\mathcal{L}_m$, can be written in terms of the superpotential as
\begin{eqnarray}\label{06}
\rho(y)=e^{2A}\Bigg[3c_{1}\alpha^{2}W^{2}+\frac{9}{4}c_{1}^{2}\alpha^{2}W_{\phi}^{2}\Bigg].
\end{eqnarray}

Now that we have set up our $f(T,B)$-braneworld model, let us examine the modifications to the brane core and source (scalar field). For this purpose, we consider two types of superpotentials. The first is a sine-Gordon type superpotential \cite{Gremm1999, Skenderis:1999mm, DeWolfe:1999cp}:
\begin{eqnarray}
W(\phi)=\beta^2\sin\Big(\frac{\phi}{\beta}\Big).\label{Wsine}
\end{eqnarray}
The second one is the deformed superpotential written as \cite{Bazeia:2003qt, Bazeia:2003aw, Bazeia:2002xg,  Cruz:2010zz}
\begin{eqnarray}
W_\beta(\phi)=\frac{\beta}{2\beta-1}\phi^{\frac{2\beta-1}{\beta}}-\frac{\beta}{2\beta+1}\phi^{\frac{2\beta+1}{\beta}},\label{Wdef}
\end{eqnarray}
where the parameter $\beta=1,\ 3,\ 5,...$ is an odd integer. The function $W_\beta(\phi)$ was originally introduced in Ref. \cite{Bazeia:2003qt} for flat spacetime. It was inspired by the work on deformed defects \cite{Bazeia:2002xg} and was obtained using a deformation procedure proposed in that work. In this procedure, the superpotential is deformed such that double-kink solutions are generated. To understand the process, let us write the superpotential associated with the standard $\phi^{4}$ models as 
\begin{equation}
    \widetilde{W}(\phi)=\phi-\phi^3/3,
\end{equation}from which one obtains the potential in flat spacetime:
\begin{equation}
    \widetilde{V}(\phi)=\frac{1}{2}(1-\phi^2)^2.
\end{equation} In Refs. \cite{Bazeia:2002xg,Bazeia:2003qt}, the deformation procedure was applied by introducing the function $f(\phi)=\phi^{1/\beta}$, where the parameter $\beta=1,\ 3,\ 5,\ \ldots$ is an odd integer. The deformed superpotential $\overline{W}\beta(\phi)$ can be obtained directly from $\widetilde{W}(\phi)$ through the relation:
\begin{equation}
\frac{df}{d\phi}\frac{d \overline{W}\beta}{d\phi}=\frac{d \widetilde{W}}{d\phi}[\phi\rightarrow f(\phi)],
\end{equation}
where in the right-hand side, we replace $\phi$ with $f(\phi)$ after taking the derivative of $\widetilde{W}$. Finally, we write $W_\beta(\phi)=1/\beta,\overline{W}_\beta(\phi)$. This superpotential was studied in the context of thick branes in Ref. \cite{Bazeia:2003aw}, where it was shown to exhibit a richer internal structure and the splitting brane effect.

\subsubsection{Sine-Gordon superpotential}

For the sine-Gordon superpotential (\ref{Wsine}), the equations (\ref{07}) and (\ref{08}) take the form
\begin{eqnarray}
\phi(y)&=&-\beta\ \mathrm{arcsin}\Big[\tanh\Big(\frac{3c_1\alpha y}{2}\Big)\Big],\\
V(\phi)&=&\frac{3}{8}c_1\alpha^2\beta^2\Big[3c_1 \cos^2\Big(\frac{\phi}{\beta}\Big)+8\beta^{2}\sin^2\Big(\frac{\phi}{\beta}\Big)\Big],
\end{eqnarray}
which is a simple thick brane solution. From Eq. (\ref{09a}), we obtain the form of the warp factor, i.e.,
\begin{eqnarray}\label{09}
A(y)=\frac{2\beta^2}{3c_1}\ln\Big[\mathrm{cosh}\Big(\frac{3c_1\alpha y}{2}\Big)\Big].
\end{eqnarray}
Finally, we can write the form of the energy density for this thick brane system, which is
\begin{eqnarray}
\rho(y)=\frac{3}{4}c_1\alpha^2\beta^2\cosh\Big(\frac{3c_1\alpha y}{2}\Big)^{-2\left(1-\frac{2\beta^2}{3c_1}\right)}\Big\{3c_1-2\beta^{2}[1-\cosh(3c_1\alpha y)]\Big\}.
\end{eqnarray}

\begin{figure}
\begin{center}
\begin{tabular}{ccc}
\includegraphics[height=5cm]{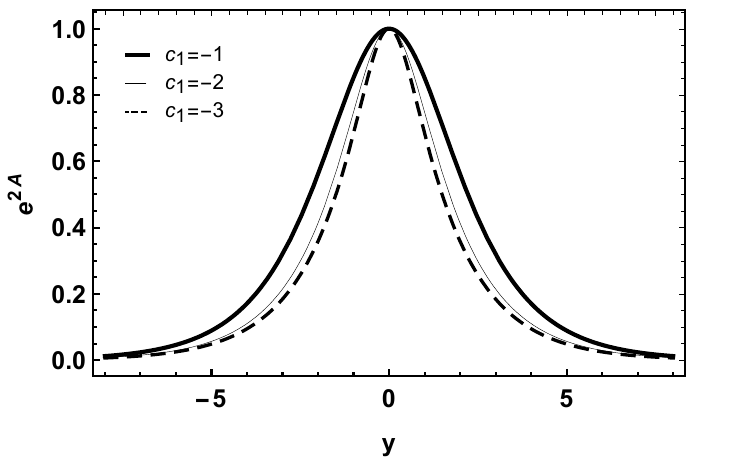}
\includegraphics[height=5cm]{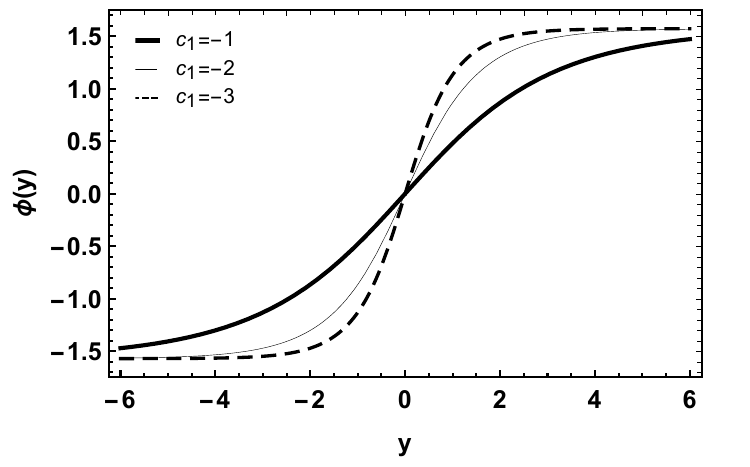}\\ 
(a) \hspace{8 cm}(b)\\
\includegraphics[height=5cm]{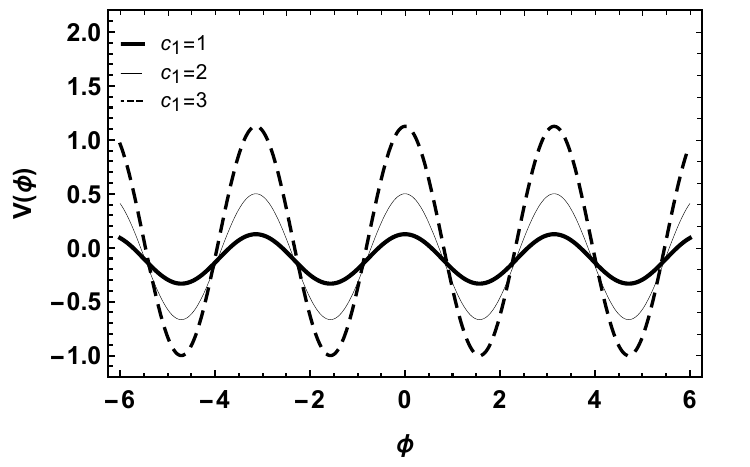}
\includegraphics[height=5cm]{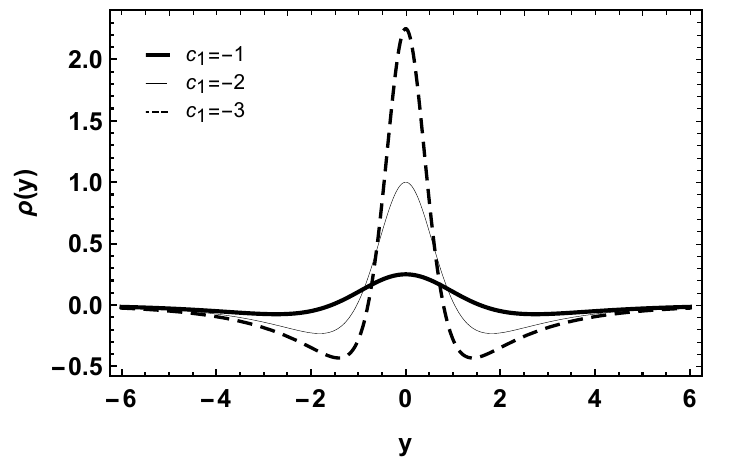}\\ 
(c) \hspace{8 cm}(d)
\end{tabular}
\end{center}
\caption{Behavior of the warp factor (a), of the kink solution (b), potential (c) and energy density (d), for the sine-Gordon type superpotential, where $\alpha=1/3$ and $\beta=1$.
\label{figene1}}
\end{figure}

Fig. \ref{figene1} shows the plots for the warp factor $e^{2A}$, scalar field $\phi$, potential $V(\phi)$, and energy density $\rho$. It is evident that the torsion parameter $c_1$ has a significant impact on the results. As $c_1$ decreases, the warp factor narrows (Fig. \ref{figene1}(a)). The scalar field has a kink-like solution, which becomes more pronounced with decreasing $c_1$ (Fig. \ref{figene1}(b)). The potential also responds to the changes in the kink-like field, resulting in modifications to its shape as $c_1$ varies (Fig. \ref{figene1}(c)). These changes directly affect the behavior of the energy density, which becomes more localized with decreasing $c_1$ (Fig. \ref{figene1}(d)).

\subsubsection{Deformed superpotential}

\begin{figure}
\begin{center}
\begin{tabular}{ccc}
\includegraphics[height=5cm]{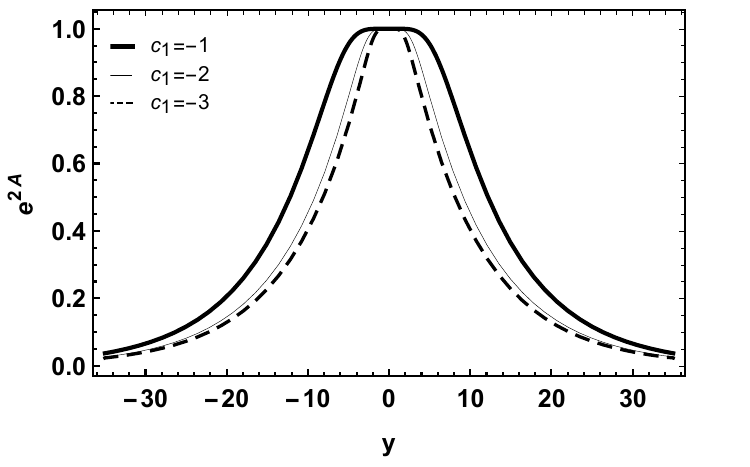}
\includegraphics[height=5cm]{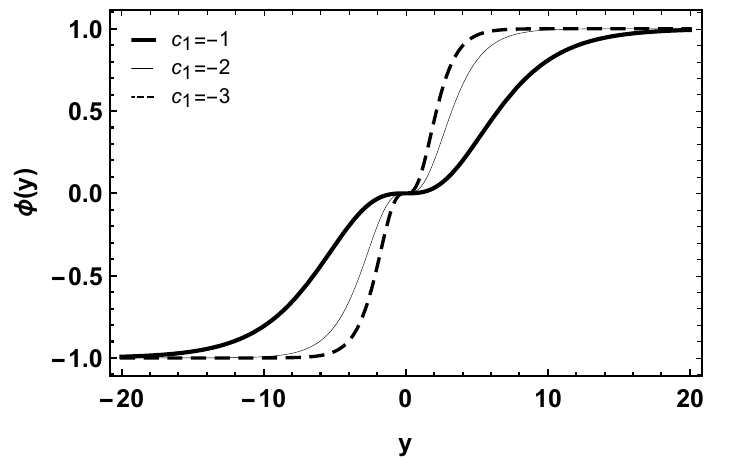}\\ 
(a) \hspace{8 cm}(b)\\
\includegraphics[height=5cm]{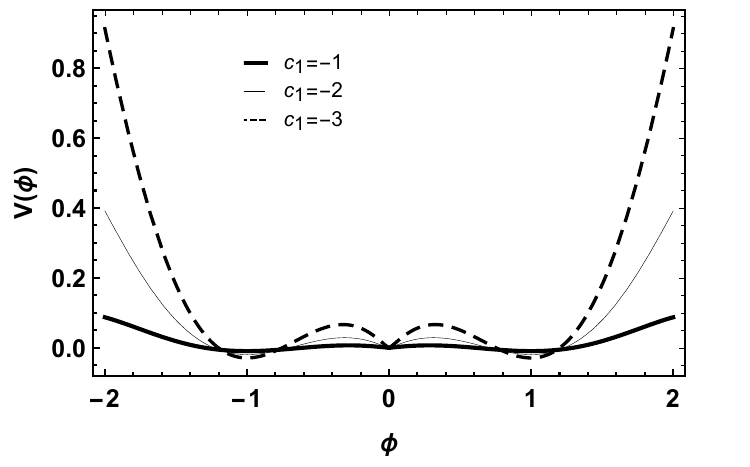}
\includegraphics[height=5cm]{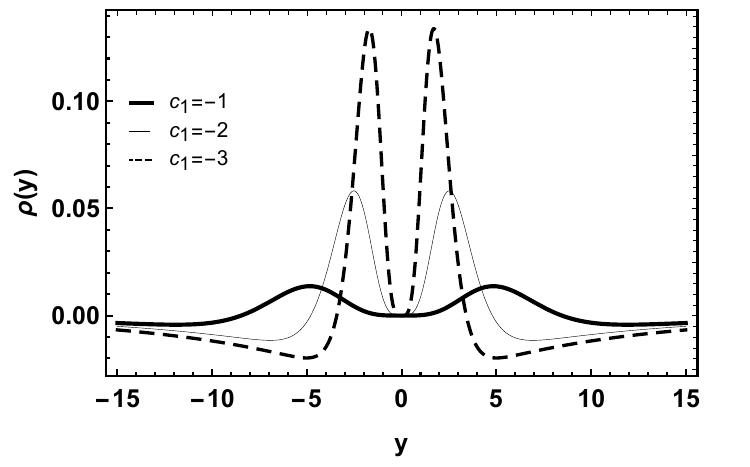}\\ 
(c) \hspace{8 cm}(d)
\end{tabular}
\end{center}
\caption{Behavior of the warp factor (a), of the kink solution (b), potential (c) and energy density (d), for the deformed superpotential, where $\alpha=1/3$ and $\beta=3$.
\label{figene2}}
\end{figure}

For the $\phi^{4}$-deformed superpotential (\ref{Wdef}), the equations (\ref{07}) and (\ref{08}) become
\begin{eqnarray}
\phi(y)&=&\left(- \tanh\Big(\frac{3c_1\alpha y}{2\beta}\Big)\right)^\beta,\\
V(\phi)&=&\frac{3}{8}c_1\alpha^2\phi^{2\left(1-\frac{1}{\beta}\right)}\Big[8\beta^2\phi^2\Big(\frac{1}{2\beta-1}-\frac{\phi^{\frac{2}{\beta}}}{2\beta+1}\Big)^2+3c_1 (\phi^{\frac{2}{\beta}}-1)^2\Big].
\end{eqnarray}
It is interesting to note that when we set $\beta=1$, we obtain the $\phi^4$ potential in a Randall-Sundrum-like scenario. However, for values of $\beta = 3,\ 5,\ 7, ...$, the potential exhibits three minima: one at $\phi = 0$, and two more at $\phi \pm 1$, as can be seen in Fig. \ref{figene2}($c$). Furthermore, the potential is also influenced by the torsion parameter, as shown by its dependence on $c_1$.

In Fig. \ref{figene2}($b$), we see how the $c_1$ parameter modifies the behavior of the $\phi$ field. For $\beta=1$ the scalar field solution is kink-like. When we consider values of $\beta=3,\ 5,\ 7, ...$ the solutions of the scalar field become double-kink-like. 

From Eq. (\ref{09a}), we can obtain the solution for the warp factor as
\begin{eqnarray}\label{aaas}
A(y)&=&\frac{\beta}{6c_1(\beta+1)(4\beta^2-1)}\Bigg[(1+3\beta+2\beta^2)(F_1^{-}+F_1^{+})\nonumber\\&-&\beta(2\beta-1)\tanh^2\Big(\frac{3c_1\alpha y}{2\beta}\Big)(F_2^{-}+F_2^{+})\Bigg]\tanh^{2\beta}\Big(\frac{3c_1\alpha y}{2\beta}\Big),
\end{eqnarray}
where 
\begin{eqnarray}
F_1^{\pm}&=&\ _2F_1\Bigg(1,2\beta,1+2\beta,\pm\tanh\Big(\frac{3c_1\alpha y}{2\beta}\Big)\Bigg),\nonumber\\
F_2^{\pm}&=& \ _2F_1\Bigg(1,2(\beta+1),3+2\beta,\pm\tanh\Big(\frac{3c_1\alpha y}{2\beta}\Big)\Bigg),
\end{eqnarray}
are the hypergeometric functions. 
For odd values $\beta>1$, we can see that the warp factor has a flattened peak. The greater the value of $\beta$, the greater the peak plateau. When we change the torsion parameter, we also change the width of the warp factor (Fig. \ref{figene2}.($a$)).

Finally, the energy density for this deformed brane is given by Eq. (\ref{aaas}):
\begin{eqnarray}
\rho(y)&=&e^{2A}\Bigg\{ 9c_{1}^{2}\alpha^{2}\left(\tanh\left(\frac{3\alpha c_{1}y}{2\beta}\right)\right)^{2\beta}\text{csch}^{2}\left(\frac{3\alpha c_{1}y}{\beta}\right) \nonumber\\
&+&3c_1\alpha^{2}\beta^{2}\left[\frac{\left(\tanh\left(\frac{3\alpha c_{1}y}{2\beta}\right)\right)^{2\beta-1}}{2\beta-1}-\frac{\left(\tanh\left(\frac{3\alpha c_{1}y}{2\beta}\right)\right)^{2\beta+1}}{2\beta+1}\right]^{2}\Bigg\}.
\end{eqnarray}

For $\beta>1$, it is possible to observe a brane split process. We can easily see from Fig. \ref{figene2}($d$) that the energy density splits into two peaks, thereby becoming more localized as the value of the parameter $c_1$ is decreased. 

\section{Gauge fields localization}
\label{sec3}

In this section, we will analyze the trapping of abelian gauge fields on brane scenarios in $f(T,B)$ gravity by considering a non-minimal Stueckelberg-like coupling between the fields and the scalar torsion and boundary term.

It is well known that five-dimensional thick brane models do not yield a normalizable zero mode for both gauge and Kalb-Ramond fields when taking the standard action for these fields. We can illustrate this problem by considering the following five-dimensional standard action for gauge vector field
\begin{equation}\label{af}
  S_{A}=-\frac{1}{4} \int d^5x h F_{MN}F^{MN},
\end{equation}
where $F_{MN}=\partial_M A_N-\partial_N A_M$ is the field strength. Now, employing Kaluza-Klein (KK) decomposition, $A_{\nu}(x,y)=\sum a_{\nu}(x)\chi(y)$, the action (\ref{af}) is rewritten as
\begin{equation}\label{af1}
 S_{A}=-\frac{1}{4}\int dy \chi(y)^2 \int d^4x  f_{\mu\nu}(x)f^{\mu\nu}(x). 
\end{equation}
with $f_{\mu\nu}=\partial_\mu a_\nu-\partial_\nu a_\mu$.

In the above expression, $x$ represents the coordinates of our universe, and $y$ represents the extra-dimension coordinate. However, we can see that the effective action (\ref{af1}) diverges due to the absence of a warp factor. This means that the gauge field zero mode is not localized on the brane. To address this issue, various coupling functions that depend on the extra dimension have been proposed in the literature \cite{Chumbes:2011zt,Zhao:2014gka}. For instance, let us assume the following action for gauge vector field
\begin{equation}\label{S3}
  S_{A}=-\frac{1}{4} \int d^5x h G(y) F_{MN}F^{MN}.
\end{equation}
Applying again the KK decomposition we obtain 
\begin{equation}\label{S33}
 S_{A}=-\frac{1}{4}\int dy G(y)\chi(y)^2 \int d^4x  f_{\mu\nu}f^{\mu\nu}.
\end{equation}
From Eq. (\ref{S33}), we write the condition for localization of gauge field as being
\begin{equation}
I=\int dy G(y)\chi(y)^2< \infty.
\end{equation}

The function $G(y)$ can be a function of the background scalar field, as it was studied in \cite{Chumbes:2011zt}, the dilaton scalar field \cite{Cruz:2010zz, Cruz:2012kd,Cruz:2013zka,Cruz:2015nrd}, or a function of the curvature scalar proposed in \cite{Zhao:2017epp}. On the other hand, in the context of modified teleparallel gravity, it can be adopted as a function of the scalar torsion as it was studied in Ref \cite{Belchior}. It is also possible to admit a mechanism of localization by adding to the action (\ref{af}) a Proca-like term given by
\begin{equation}
 S_{PL}=-\frac{1}{2}\int d^5x hG(y)A_M A^M,  \label{massProca}
\end{equation}where the function $G(y)$ plays the role of a dynamical mass for gauge vector field. This mechanism was studied in \cite{Zhao:2014iqa} by considering $G(y)=-1/16 R$. However, the action (\ref{massProca}) has the disadvantage of breaking the 5D gauge symmetry. In this work, we are interested in exploring an alternative method studied by Vaquera and Corradini in Ref. \cite{Vaquera-Araujo:2014tia}, where a Stueckelberg-like action is assumed as
\begin{equation}
 S_{SL}=-\frac{1}{2}\int d^5x hG(y)(A_M-\partial_M S)^2,  
\end{equation}
where $S$ is the Stueckelberg-like scalar field. This action has the advantage of being gauge-invariant. It is important to point out that the Stueckelberg-like field acts as an auxiliary field and does not represent a physical degree of freedom, and it will not affect either the zero-mode or the massive modes as it will be shown ahead. In this sense, we introduced the Stueckelberg-like scalar field only to obtain a gauge-invariant mechanism of field localization. Moreover, such a mechanism can be straightforwardly extended to the Kalb-Ramond field \cite{Vaquera-Araujo:2014tia}.  

Now, our goal is to use this kind of coupling to investigate the localization of the gauge vector and Kalb-Ramond fields on the braneworld scenarios constructed in $f(T,B)$ gravity. To this end, we will consider a function $G$ that depends on the torsion scalar and the boundary term.

\subsection{Gauge vector field}
Based on Ref. \cite{Vaquera-Araujo:2014tia}, we define a five-dimensional Stueckelberg-like action for gauge vector field as follows:
\begin{equation}\label{667}
 S_{A-SL}=\int d^5x h\Big[-\frac{1}{4}F_{MN}F^{MN}-\frac{1}{2}G(T,B)(A_M-\partial_M S)^2\Big],   
\end{equation}
where $G(T,B)$ is a suitable function of the torsion scalar and the boundary term. Later, we will adopt a particular choice for this function, which is able to produce a normalizable zero mode for the gauge vector field. It is worth mentioning that action (\ref{667}) is invariant under the five-dimensional gauge transformations
\begin{eqnarray}
A_M&\rightarrow& A_M+\partial_M \Lambda,\nonumber\\
S&\rightarrow& S+\Lambda.
\end{eqnarray}

Now, we vary the action (\ref{667}) to obtain the following equation of motion for $A_M$ and $S$
\begin{eqnarray}
 \partial_M\Big(e^{4A}g^{ML}g^{NP}F_{LP}\Big)&=&- e^{4A}G(T,B)g^{NP}(A_P-\partial_P S),\nonumber\\
 \partial_M\Big(e^{4A}G(T,B)g^{ML}(A_L-\partial_L S)\Big)&=&0.  
\end{eqnarray}
The field $A_M$ can be parameterized as
\begin{equation}
 A_M=(A_\mu,A_4)=(\widehat{A}_\mu+\partial_\mu\phi,A_4), \end{equation}
being $\widehat{A}_\mu$ the transverse component of $A_\mu$ with the condition $\partial_\mu \widehat{A}^\mu=0 $, while $\partial_\mu\phi$ longitudinal component. These components transform under four-dimensional gauge transformation as follows
\begin{eqnarray}
 \widehat{A}_\mu &\rightarrow& \widehat{A}_\mu,\nonumber\\
 A_4 &\rightarrow&  A_4 + \Lambda^{\prime},\nonumber\\
 S &\rightarrow& S + \Lambda,\nonumber\\
 \phi&\rightarrow& \phi + \Lambda.
\end{eqnarray}
With these transformation, it is convenient to redefine  $A_4$ and $S$ as
\begin{eqnarray}
A_4 &\rightarrow& \lambda + \phi^{\prime},\nonumber\\
S &\rightarrow& \rho + \phi.
\end{eqnarray}
The above redefinition it is useful since the new fields are invariant. Now, by considering this new parameterization, we obtain the following set of equations 
\begin{eqnarray}
 \label{eA}[e^{-2A}\square +\partial_4^2+2A^{\prime}\partial_4-G]\widehat{A}_{\nu}&=&0,\\
 \partial_4(e^{2A}\lambda)-e^{2A}G\rho&=&0,\\
 e^{2A}\square\lambda+e^{4A}G(\rho^{\prime}-\lambda)&=&0,\\
 e^{2A}G\square\rho+\partial_4[e^{4A}G\rho^{\prime}-\lambda)]&=&0,
\end{eqnarray}
where $\square =\eta^{\mu\nu}\partial_\mu\partial_\nu$.

Now that we have obtained the set of equations of motion, the next step is to study the localization of the transverse component $\widehat{A}{\nu}$. It is important to note that the scalar part is decoupled from the vector one. Therefore, the transverse component can be analyzed separately. With this in mind, let us adopt the Kaluza-Klein decomposition for $\widehat{A}{\nu}$ as follows:\begin{equation}
\widehat{A}_{\nu}(x,y)=\sum a_{\nu}(x)\chi(y),   
\end{equation}
with $\square a_{\nu}(x)=m^2 a_{\nu}(x)$. Thus, the equation (\ref{eA}) becomes
\begin{equation}
 \chi^{\prime\prime}+2A^{\prime} \chi^{\prime} -G\chi=-m^2 e^{-2A}\chi.
\end{equation}

By considering the conformal coordinate $dz=e^{-A}dy$, the above equation reduce to
\begin{equation}\label{898}
 \Ddot{\chi} +\Dot{A}\Dot{\chi}-Ge^{2A}\chi =-m^2\chi,
\end{equation}
where the dot indicate differentiation with respect to conformal coordinate $z$. In order to analyze the zero mode and massive mode, the equation (\ref{898}) is written in Schr\"{o}dinger-like form by means of following change $\chi(z)=e^{-\frac{A}{2}}\psi(z)$, thus, we get
\begin{equation}\label{phi}
 -\Ddot{\psi}+V_{eff}\psi=m^2\psi,   
\end{equation}
where we write the effective potential $V_{eff}$ as
\begin{equation}\label{epa}
V_{eff}=\frac{1}{4}\Dot{A}^2+\frac{1}{2}\Ddot{A}+e^{2A}G.
\end{equation}

We can now analyze the influence of scalar torsion and the boundary term on the localization of the transverse component $\widehat{A}_\mu$. To do this, we need to choose a particular form for the function $G$. Let us adopt a linear combination, i.e., $G(T,B)=\gamma_1T+\gamma_2B$. Note that when we fix $\gamma_1=c_1$ and $\gamma_2=c_2$, $G(T,B)$ has the same form as the function $f(T,B)$ defined previously in Sec. \ref{sec2}. This choice of $G$ is interesting because it allows us to recover the influence of the boundary term $B$. Furthermore, our potential takes a simple form that guarantees localized modes. With this choice, the effective potential (\ref{epa}) becomes\begin{equation}\label{pot}
 V_{eff}=\Big[\frac{1}{4}-12(\gamma_1+2\gamma_2)\Big]\Dot{A}^2+\Big(\frac{1}{2}-8\gamma_2\Big)\Ddot{A}, 
\end{equation}
which behaves like an even function. This assures us that the massive solutions of Eq. (\ref{phi}) can be even or odd. To numerically solve Eq. (\ref{phi}), let us use the following boundary conditions
\begin{eqnarray}
\varphi_{even}(0)&=&c,\ \ \partial_z\varphi_{even}(0)=0,\nonumber\\
\varphi_{odd}(0)&=&0, \ \ \partial_z\varphi_{odd}(0)=c,
\end{eqnarray}
which represent the massive modes even and odd parity \cite{Liu2009}. Here $c$ is just a constant.

The last step is to ensure a normalizable zero mode and positivity (absence of tachyon modes). We have verified that the theory remains stable because equation (\ref{phi}) can be factorized in the form \cite{Vaquera-Araujo:2014tia}:
\begin{equation}
Q_{\xi}^{\dagger}Q_{\xi}\psi = m^2\psi,
\end{equation}
where $m^{2}\geq 0$. The operators are defined as follows:
\begin{eqnarray}
Q_{\xi}^{\dagger}=\partial_z+\left(\xi+\frac{1}{2}\right)\Dot{A},\nonumber\
Q_{\xi}=-\partial_z+\left(\xi+\frac{1}{2}\right)\Dot{A},
\end{eqnarray}
where $\xi=\pm\sqrt{-12\gamma_1-16\gamma_2}$ and $\dagger$ denotes Hermitian conjugation. The massless mode is given by
\begin{equation}\label{msm}
 \psi_{0}(z)=k_{0}e^{\Big(\xi+\frac{1}{2}\Big)A(z)},  
\end{equation}
where $k_0$ is just a normalization constant. For $\gamma_1=1$ and $\gamma_2=-1$, we recover the massless mode obtained in \cite{Zhao:2014iqa}. Ahead it is plotted the behavior of massless mode $\psi_{0}(z)$ and the potential $V(z)$ for both Sine-Gordon (Fig. \ref{fig3}) and deformed superpotential (Fig. \ref{fig4}) where it will be considered $\gamma_1=c_1$ and $\gamma_2=c_2$. 

\subsubsection{Sine-Gordon superpotential}

\begin{figure}
\begin{center}
\begin{tabular}{ccc}
\includegraphics[height=5cm]{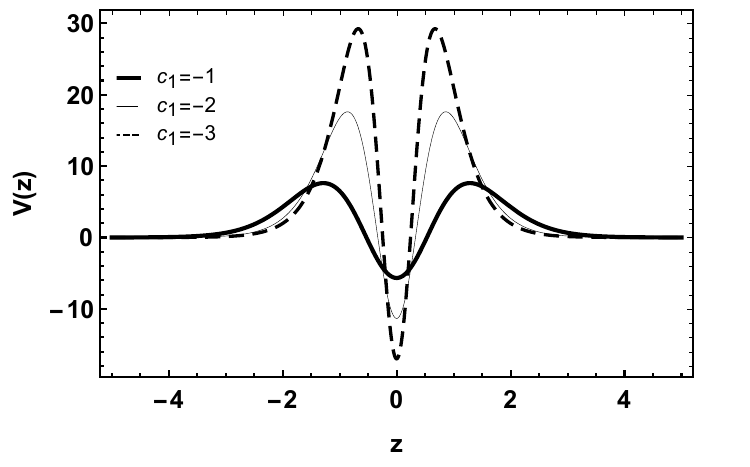}
\includegraphics[height=5cm]{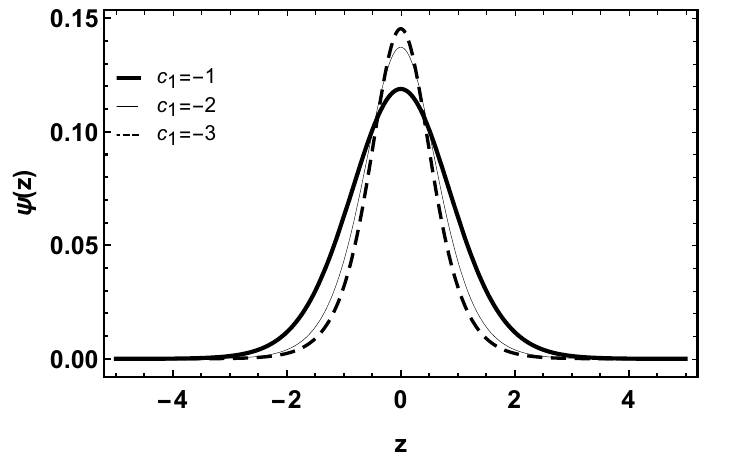}\\ 
(a) \hspace{8 cm}(b)\\
\includegraphics[height=5cm]{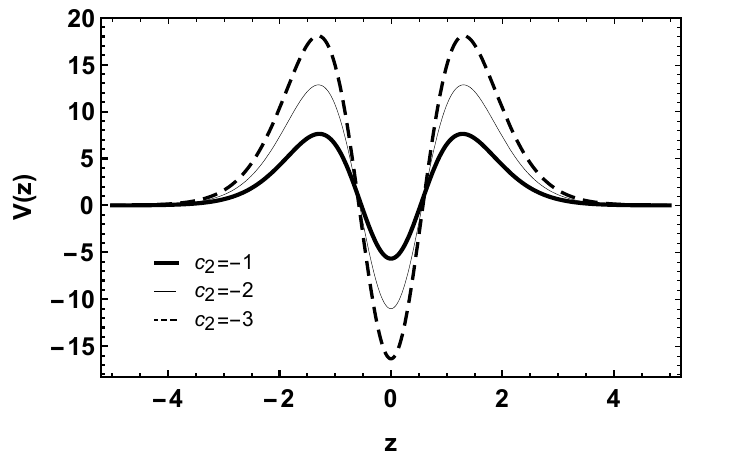}
\includegraphics[height=5cm]{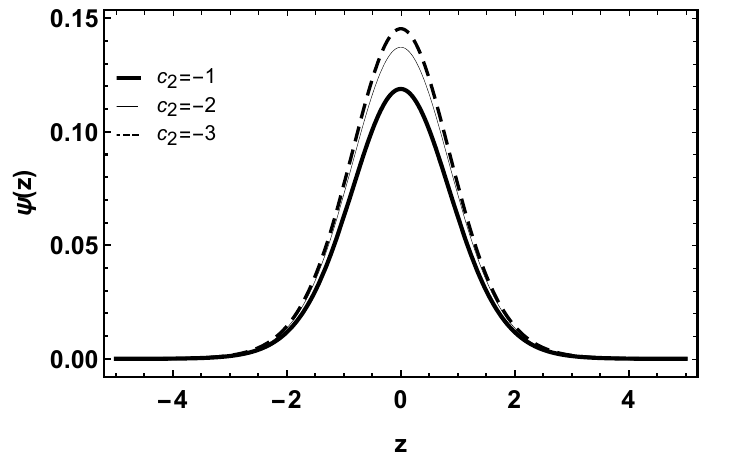}\\ 
(c) \hspace{8 cm}(d)
\end{tabular}
\end{center}
\caption{
Potential and massless mode for the sine-Gordon type superpotential, where $\alpha=1/3$ and $\beta=1$. (a) and (b) with $c_2=-1$. (c) and (d) with $c_1=-1$. 
\label{fig3}}
\end{figure}

 The influence of torsion and boundary term (by varying the parameters $c_1$ and $c_2$ respectively) on the potential and its direct influence on massless modes can be analyzed. As the value of the torsion parameter decreases, the potential increases its maximum (peaks and valleys) making the massless modes more localized (Fig.\ref{fig3}.($a$) and ($b$)). A similar effect is seen when we decrease the value of the boundary term parameter (Fig.\ref{fig3}.($c$) and ($d$)).

\begin{figure}
\begin{center}
\begin{tabular}{ccccccccc}
\includegraphics[height=5cm]{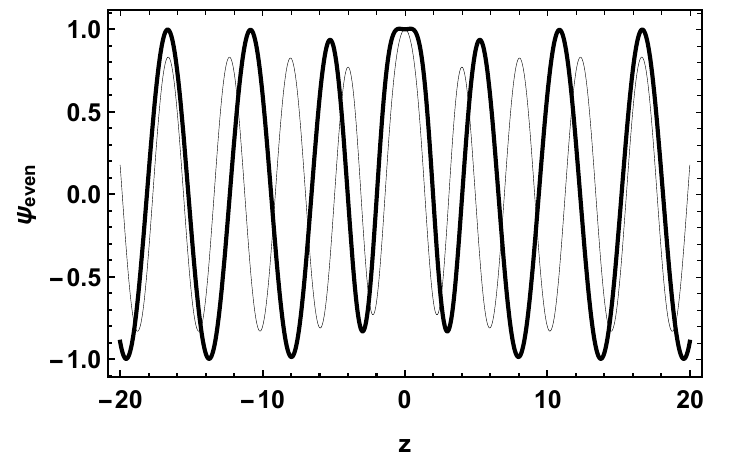}
\includegraphics[height=5cm]{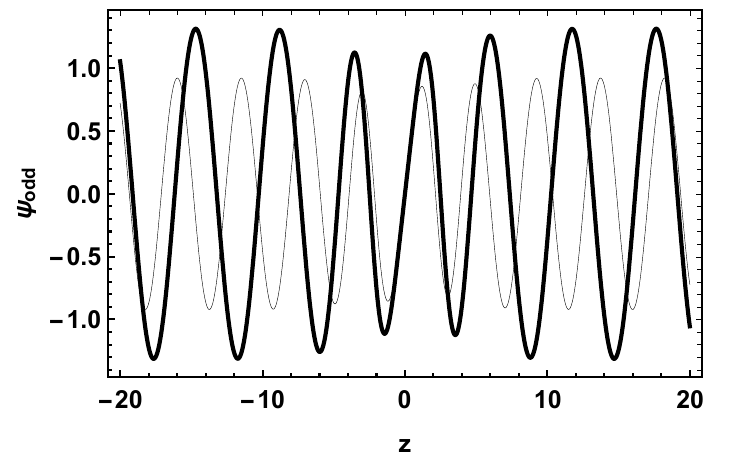}  \\
(a)\hspace{8 cm}(b)
\end{tabular}
\end{center}
\caption{Massive modes for the sine-Gordon type superpotential, where $\alpha=1/3$ and $\beta=c_1=c_2=-1$. (a) Even solution with thick line $m=1.08$ and thin line $m=1.46$. (b) Odd solution with thick line $m=1.06$ and thin line $m=1.40$.} 
\label{fig5}
\end{figure}

\begin{figure}
\begin{center}
\begin{tabular}{ccc}
\includegraphics[height=5cm]{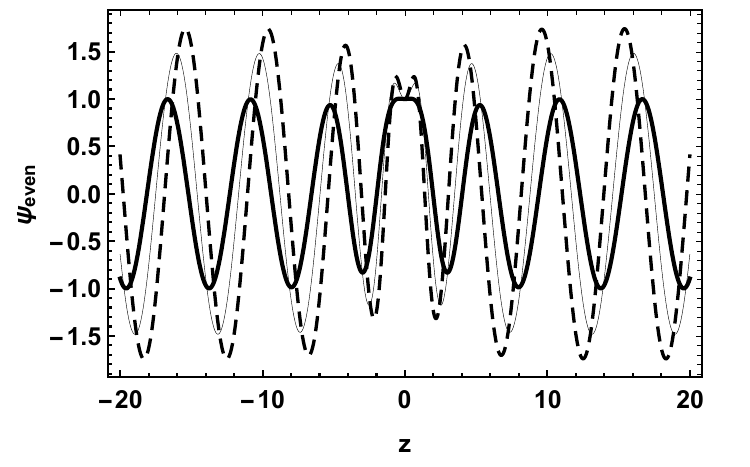}
\includegraphics[height=5cm]{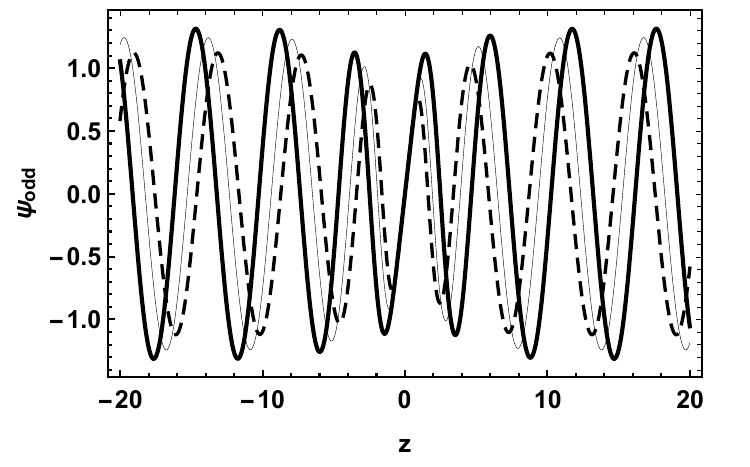}\\ 
(a) \hspace{8 cm}(b)\\
\includegraphics[height=5cm]{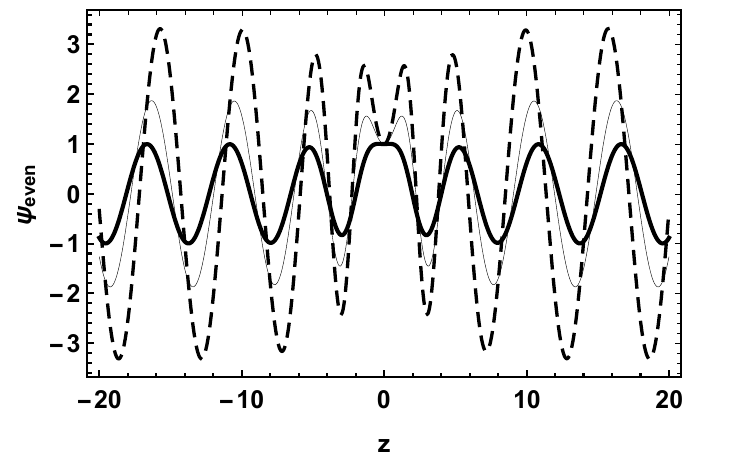}
\includegraphics[height=5cm]{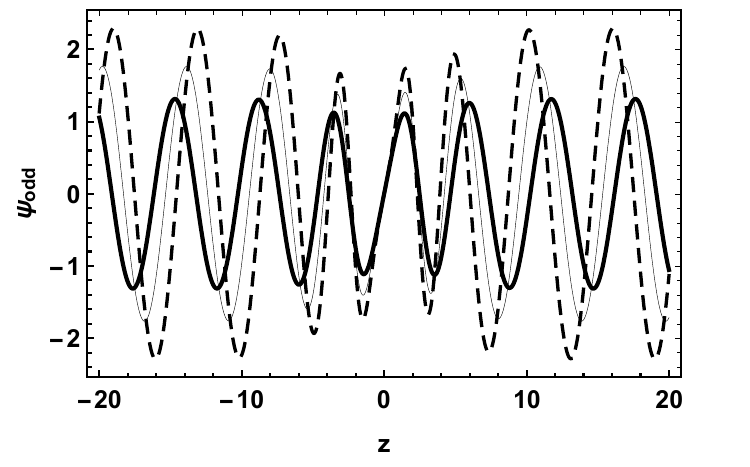}\\ 
(c) \hspace{8 cm}(d)
\end{tabular}
\end{center}
\caption{Massive modes for the sine-Gordon type superpotential, where $\alpha=-1/3$ and $\beta=1$. Being $c_2=-1$ and with the thick line $c_1=-1$, thin line $c_1=-2$ and dashed line $c_1=-3$, even solution (a) and odd solution (b). Being $c_1=-1$ and with the thick line $c_2=-1$, thin line $c_2=-2$ and dashed line $c_2=-3$, even solution (c) and odd solution (d). For even solution $m=1.08$ and odd solution $m=1.06$.
\label{fig6}}
\end{figure}

Figures \ref{fig5} and \ref{fig6} show the numerical solutions of Eq. (\ref{phi}). In Fig. \ref{fig5}, we observe the change in the solutions as the mass eigenvalues are modified. Increasing the mass eigenvalues causes the solutions to have more oscillations and lower amplitudes. However, when we fix the mass value and vary the torsion parameters and boundary term, a more interesting behavior emerges. Specifically, decreasing the value of the torsion parameter causes the even and odd solutions to move closer to the brane core and increase in amplitude (Fig. \ref{fig6}.($a$) and ($b$)). Similarly, decreasing the value of the boundary term parameter leads to similar behavior in the solutions (Fig. \ref{fig6}.($c$) and ($d$)).

\subsubsection{Deformed superpotential}

\begin{figure}
\begin{center}
\begin{tabular}{ccc}
\includegraphics[height=5cm]{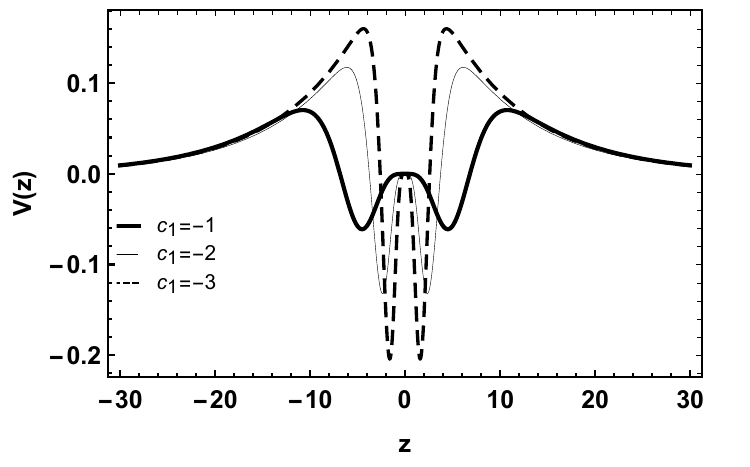}
\includegraphics[height=5cm]{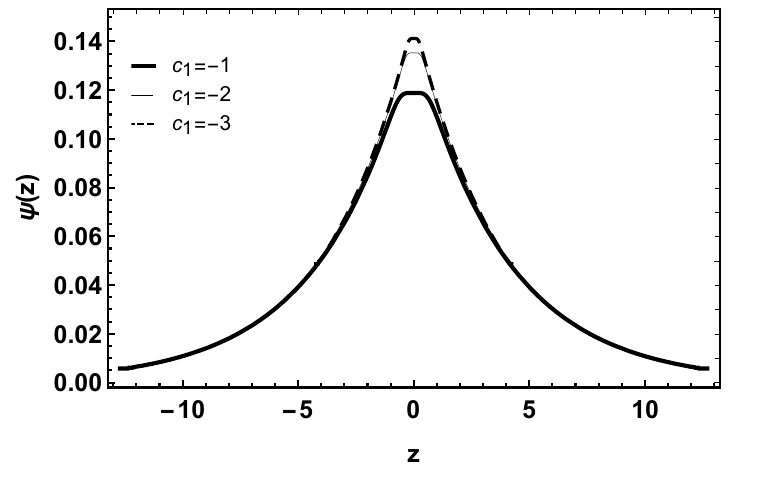}\\ 
(a) \hspace{8 cm}(b)\\
\includegraphics[height=5cm]{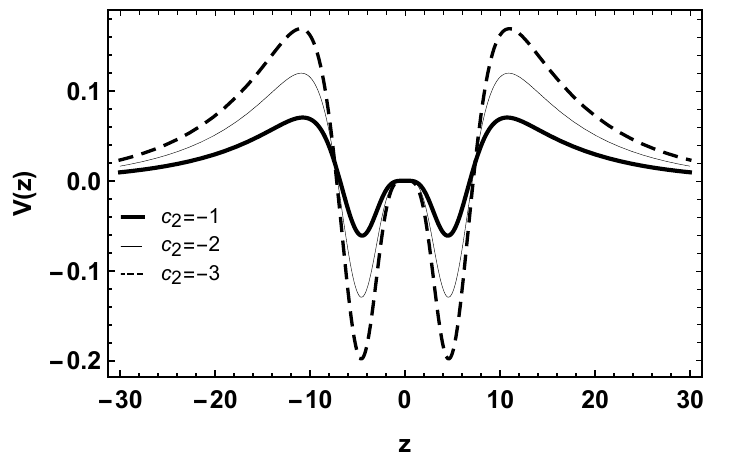}
\includegraphics[height=5cm]{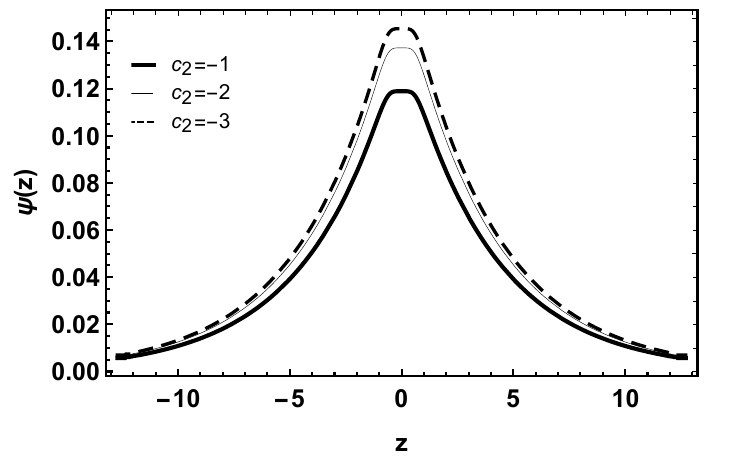}\\ 
(c) \hspace{8 cm}(d)
\end{tabular}
\end{center}
\caption{
Potential and massless mode for the deformed superpotential, where $\alpha=1/3$ and $\beta=3$. (a) and (b) with $c_2=-1$. (c) and (d) with $c_1=-1$. 
\label{fig4}}
\end{figure}

For the deformed brane, we obtain a potential that represents the solutions obtained in section \ref{sec2}, where the brane tries to split. The potential has an internal structure with minimums which suffer the influence of parameters $c_{1,2}$. When we vary the values of the torsion parameters and the boundary term, these minimums tend to become more accentuated (Fig. \ref{fig4}.($a$) and ($c$)). Massless modes feel the change in potential, tending to become more localized as we decrease the parameters $c_1$ and $c_2$ (Fig. \ref{fig6}.($b$) and ($d$)).

\begin{figure}
\begin{center}
\begin{tabular}{ccc}
\includegraphics[height=5cm]{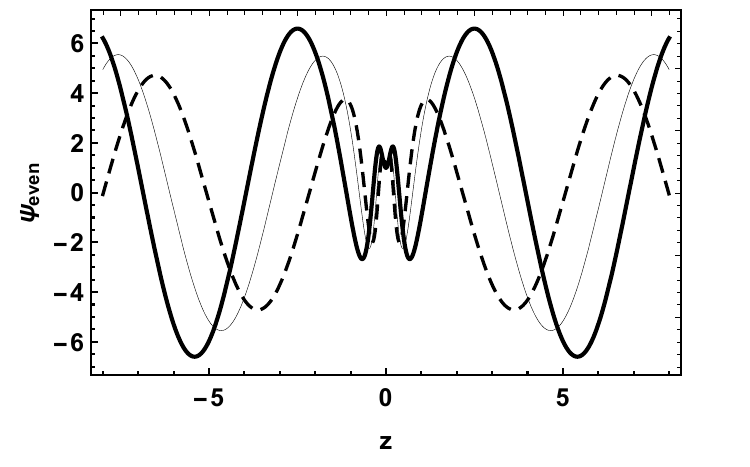}
\includegraphics[height=5cm]{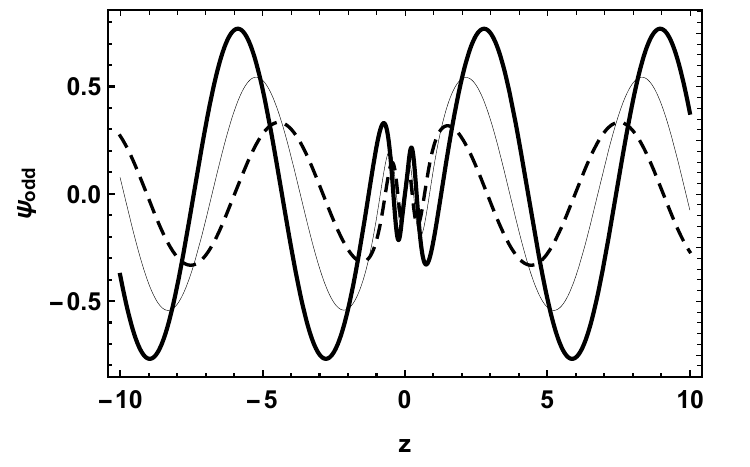}\\ 
(a) \hspace{8 cm}(b)\\
\includegraphics[height=5cm]{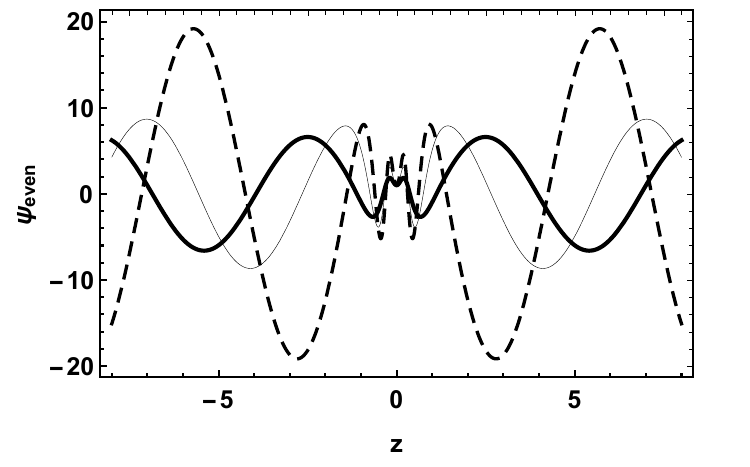}
\includegraphics[height=5cm]{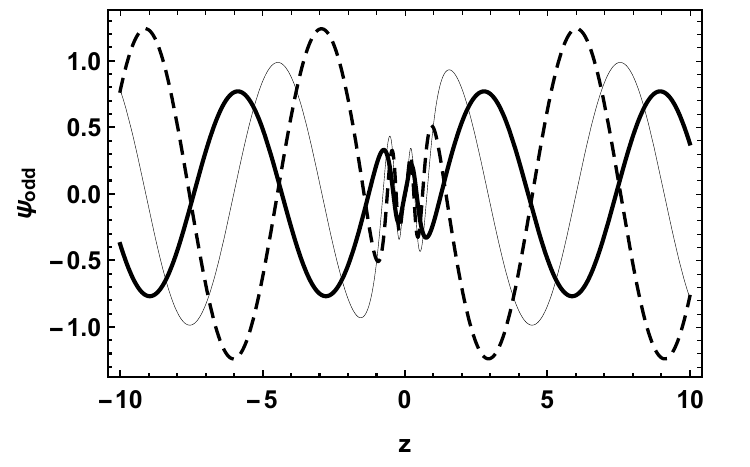}\\ 
(c) \hspace{8 cm}(d)
\end{tabular}
\end{center}
\caption{Massive modes for the deformed superpotential, where $\alpha=1/3$ and $\beta=3$. Being $c_2=-1$ and with the thick line $c_1=-1$, thin line $c_1=-2$ and dashed line $c_1=-3$, even solution (a) and odd solution (b). Being $c_1=-1$ and with the thick line $c_2=-1$, thin line $c_2=-2$ and dashed line $c_2=-3$, even solution (c) and odd solution (d). For even solution $m=1.40$ and odd solution $m=1.02$.
\label{fig7}}
\end{figure}

Figure \ref{fig7} illustrates the behavior of massive modes by varying the parameters that control the influence of torsion and boundary term. When decreasing the value of the torsion parameter, we decrease the amplitudes that are shifted closer to the brane (Fig. \ref{fig7}.($a$) and ($b$)). On the other hand, when we decrease the boundary term values, the oscillation amplitudes increase (Fig. \ref{fig7}.($c$) and ($d$)). A similar behavior was seen in Ref. \cite{Cruz:2010zz}, where the gauge field localization was addressed in the context of general relativity by considering a gauge-invariant coupling with the dilaton field. In our case, we show that only the torsion scalar and boundary term are enough to gauge field zero-mode trapping. As we pointed out before, although we have assumed a Stueckelberg-like field to obtain a gauge-invariant action, we could decouple an equation only for transverse component $\widehat{A}_\mu$ that allows our analysis. In this sense, the Stueckelberg-like field works as an auxiliary field, thereby not contributing directly to gauge field zero-mode.

\subsection{Kalb-Ramond field}
In order to study the trapping of the Kalb-Ramond field on thick brane, we can use an analogous procedure to that discussed in the previous section. Thus, the action can be written as \cite{Vaquera-Araujo:2014tia}
\begin{equation}\label{akr}
 S_{B-SL}=\int d^5x h\Big[-\frac{1}{6}H_{LMN}H^{LMN}-\frac{1}{2}G(T,B)(B_{MN}-C_{MN})^2\Big].   
\end{equation}
where $H_{LMN}=\partial_L B_{MN}+\partial_M B_{NL}+\partial_N B_{LM}$ and $C_{MN}=\partial_M C_N-\partial_N C_M$, being $C_M$ the Stueckelberg vector field. With the introduction of Stueckelberg-like field, the action (\ref{akr}) becomes invariant under the following $5D$ gauge transformation
\begin{eqnarray}
 B_{MN}&\rightarrow& B_{MN}+\partial_M\Lambda_N-\partial_N\Lambda_M,\nonumber\\
 C_M&\rightarrow& C_M+\Lambda_M.
\end{eqnarray}

On the another hand, the equation for $B_{MN}$ and $C_{MN}$ are written as
\begin{eqnarray}
 \partial_M\Big(e^{4A}g^{MQ}g^{NR}g^{LS}H_{QRS}\Big)&=&-e^{4A}g^{NR}g^{LS}(C_{RS}-B_{RS}),\nonumber\\
 \partial_M\Big(e^{4A}g^{MQ}g^{NR}(C_{QR}-B_{QR})\Big)&=&0.
\end{eqnarray}
By following the same procedure as previous section, the field $B_{MN}$ is parameterized as
\begin{equation}
B_{MN}=\begin{pmatrix} 
\widehat{B}_{\mu\nu}+\partial_\mu\phi_\nu-\partial_\nu\phi_\mu & B_{4\rho}\\
B_{\rho 4} &0
\end{pmatrix},    
\end{equation}
where $\widehat{B}_{\mu\nu}$ represents the transverse component of $B_{\mu\nu}$ and $\phi_\mu$ is vector such that $\partial_\lambda\phi_{\mu\nu}+\partial_\mu\phi_{\nu\lambda}+\partial_\nu\phi_{\lambda\mu}=0$ and $\phi_{\mu\nu}=\partial_\mu\phi_\nu-\partial_\nu\phi_\mu$. The $4D$ gauge transformations are
\begin{eqnarray}
\widehat{B}_{\mu\nu}&\rightarrow& \widehat{B}_{\mu\nu},\nonumber\\
\phi_\mu&\rightarrow&\phi_\mu +\lambda_\mu,\nonumber\\
B_{4\mu}&\rightarrow& B_{4\mu}+\partial_4\Lambda_\mu-\partial_\mu\Lambda_4.
\end{eqnarray}
As in the previous section, it is useful to use the following definitions
\begin{eqnarray}
 \lambda_\mu&=&B_{4\mu}-\phi_\mu+\partial_\mu C_4,\nonumber\\
 \lambda_{\mu\nu}&=&\partial_\mu\lambda_\nu-\partial_\nu\lambda_\mu,\nonumber\\
 \rho_\mu&=&C_\mu-\phi_\mu,\nonumber\\
 \rho_{\mu\nu}&=&\partial_\mu \rho_\nu-\partial_\nu \rho_\mu.
\end{eqnarray}
With these considerations, we get
\begin{eqnarray}
 \label{eB}[e^{-2A}\square+(\partial_4^2-G)]B_{\mu\nu}&=&0,\\
 \lambda_\mu^{\prime}-G\rho_\mu&=&0,\\
 \partial^\mu\lambda_{\mu\nu}+e^{2A}G(\rho^{\prime}-\lambda)&=&0,\\
 G\partial^\mu\rho_{\mu\nu}+\partial_4[e^{2A}(\rho-\lambda)]&=&0.
\end{eqnarray}

Similarly to the gauge field case, it is possible to write the transverse component $\widehat{B}_{\mu\nu}$ with the vector part isolatedly. The KK decomposition for $\widehat{B}_{\mu\nu}$ is given by
\begin{equation}\label{B}
\widehat{B}_{\mu\nu}(x,y)=\sum b_{\mu\nu}(x)\chi(y).    
\end{equation}
Substituting (\ref{B}) into (\ref{eB}), we get
\begin{equation}\label{090}
 \chi^{\prime\prime} -G\chi=-m^2 e^{-2A}\chi.
\end{equation}
Again, we use the conformal coordinate $dz=e^{-A}dy$ to write the equation (\ref{090}) as
\begin{equation}
 \Ddot{\chi} -\Dot{A}\Dot{\chi}-Ge^{2A}\chi =-m^2\chi,
\end{equation}
which can be put into a Schr\"{o}dinger-like equation through the change $\chi(z)=e^{\frac{A}{2}}\psi(z)$, thereby becoming
\begin{equation}
 -\Ddot{\psi}+V_{eff}\psi=m^2\psi,  
\end{equation}
with the effective potential being
\begin{equation}\label{vkr}
V_{eff}=\frac{1}{4}\Dot{A}^2-\frac{1}{2}\Ddot{A}+e^{2A}G.
\end{equation}

Using the same $G(T,B)$ function as in the previous section, i.e., $G(T,B)=\gamma_1T+\gamma_2B$, the potential (\ref{vkr}) takes the form
\begin{equation}\label{pot2}
  V_{eff}=\Big[\frac{1}{4}-12(\gamma_1+2\gamma_2)\Big]\Dot{A}^2+\Big(-\frac{1}{2}-8\gamma_2\Big)\Ddot{A}.   
\end{equation}
Again, we use the operators defined in the previous section to show that there are no tachyon modes and the normalizable zero mode is confined on $f(T,B)$-thick brane, being written as
\begin{equation}\label{msy}
 \psi_{0}(z)=k_{0}e^{\Big(\xi+\frac{1}{2}\Big)A(z)},  
\end{equation}
where $\xi=\pm\sqrt{\frac{1}{4}-12\gamma_1-16\gamma_2}$ and $k_0$ is a normalization constant.

At this point, it is worth observing that the only difference between the potential obtained for the gauge vector field (\ref{pot}) and the potential obtained for the KR field (\ref{pot2}) is only one sign in a constant. Therefore, the behavior of the solutions (massive and massless modes) for the KR field are similar to those already demonstrated for the gauge field. In this sense, it would be repetitive to represent such solutions graphically.

\section{Final remarks}\label{sec4}

In this work, we have considered thick braneworld models in the context of modified teleparallel theories. We first introduced the basic concepts of teleparallel gravity, which are fundamentally different from those of general relativity. To construct a thick brane scenario in $f(T,B)$ gravity, we used a Randall-Sundrum-like line element and a standard single scalar field. Employing the first-order formalism, we considered two superpotential models, sine-Gordon and $\phi^{4}$-deformed, and a linear form for the function $f(T,B)=c_1T+c_2B$, where the parameters $c_{1,2}$ represent a possible extension of usual teleparallelism. As we showed, the warp factor depends on these parameters.

As we have pointed out, abelian gauge fields are not confined on the brane when a standard action is taken. In this sense, a suitable coupling is required to produce a normalizable zero mode. Various types of coupling have been proposed in the literature to address this issue. In this work, we introduce a new localization mechanism to analyze the trapping of the gauge vector and the Kalb-Ramond fields in the context of $f(T,B)$ gravity. Specifically, we introduce a Stueckelberg-like interaction to non-minimally couple the fields to the torsion scalar and boundary term.

In the sequel, we have studied a mechanism based on Stueckelberg-like geometrical coupling that supports the trapping of massless mode for the transverse component of the gauge vector and the Kalb-Ramond fields. These couplings allow us to analyze the influence of torsion scalar and boundary term on the localization of these fields. The parameters $c_{1,2}$ play an important role since they control the localization of massless and massive modes and directly affect the effective potential $V_{eff}$. Our results generalize those obtained in \cite{Cruz:2010zz} and \cite{Cruz:2015nrd} with a dilaton coupling in the context of general relativity. Moreover, a remarkable fact about our coupling is the absence of massive tachyonic modes for both fields, where we have used the Schr\"{o}dinger approach. All these conditions demonstrate the consistency of our $f(T,B)$-thick brane models.

\section*{Acknowledgments}
\hspace{0.5cm} The authors thank the Funda\c{c}\~{a}o Cearense de Apoio ao Desenvolvimento Cient\'{i}fico e Tecnol\'{o}gico (FUNCAP), the Coordena\c{c}\~{a}o de Aperfei\c{c}oamento de Pessoal de N\'{i}vel Superior (CAPES), and the Conselho Nacional de Desenvolvimento Cient\'{i}fico e Tecnol\'{o}gico (CNPq), Grants no. 200879/2022-7 (RVM) and no. 309553/2021-0 (CASA) for financial support. R. V. Maluf acknowledges the Departament de F\'{i}sica Te\`{o}rica de la Universitat de Val\`{e}ncia for the kind hospitality. The authors also thank the anonymous referee for their valuable comments and suggestions.


\end{document}